# A nexus between 3D atomistic data hybrids derived from atom probe microscopy and computational materials science: a new analysis of solute clustering in Al-alloys

Baptiste Gault[1], Xiang Yuan Cui[2], Michael P. Moody[3], Anna V. Ceguerra[2], Andrew J. Breen[1,2], Ross K.W. Marceau[4], Simon P. Ringer[2]

AFFILIATIONS:

[1] Max-Planck-Institut für Eisenforschung, Düsseldorf, Germany

[2] Australian Institute for Nanoscale Science and Technology, and School of Aerospace, Mechanical and Mechatronic Engineering, The University of Sydney, NSW, 2006, Australia

[3] Department of Materials, University of Oxford, Parks Road, Oxford, OX1 3PH, United Kingdom.

[4] Deakin University, Institute for Frontier Materials, Geelong, VIC, 3216, Australia.

**Abstract:**

Solute clusters affect the physical properties of alloys. Knowledge of the atomic structure of solute clusters is a prerequisite for material optimisation. In this study, solute clusters in a rapid-hardening Al-Cu-Mg alloy were characterised by a combination of atom probe tomography and density functional theory, making use of a hybrid data type that combines lattice rectification and data completion to directly input experimental data into atomistic simulations. The clusters input to the atomistic simulations are thus observed experimentally, reducing the number of possible configurations. Our results show that spheroidal, compact clusters are more energetically favourable and more abundant.

**Introduction**

Solute clusters form during the early stages of the decomposition of most supersaturated solid solutions. The clustering of solutes is the resultant balance between complex solute-solute and solute-vacancy chemical bonding preferences in multicomponent materials systems. These processes can be subtle or acute, transient or sustained, and the result can be the formation of a state of clustering in the solid solution where the average solute atom clusters may contain just a few atoms. Segregation and clustering are not to be confused as the same thing: segregation is the partitioning, or coalescence of solute to a particular site, usually a defect such as a dislocation, grain boundary, surface or interphase interface. Clustering, on the other hand, is the discrete phenomena of local aggregation of solute atoms such that the solute clusters are uniformly dispersed. Atomic clusters are known to have significant effects on the mechanical [1–6] and other physical [7,8] properties of the material as well as on the transformation pathways of the material upon further ageing or over its lifetime[9]. Deploying materials design approaches that exploit such clustering processes, or hinder their formation, requires a knowledge of the precise energetics of these atomic-scale microstructural features, and this can be gained from atomistic simulations.

Recent work by Stephenson et al. [10] aimed to quantify the number of possible configurations of clusters defined by a succession of atoms in first nearest-neighbour positions, as well as the likelihood to be found in a given alloy system as a function of the alloy composition and short-range order parameter. The results show that there are over $10^9$ configurations for a cluster of 10 solute atoms of a single species in a face-centred cubic matrix. Mixing several solutes further adds to the complexity. This complexity is what led Vaithyanathan and co-workers to indicate that the configurational entropy associated with the disorder, makes calculating the free energy of a random solid solution almost inaccessible via direct first principles simulations [11]. Specific strategies, such as cluster-expansion, make the assessment of the free energy possible [12–15], but they rely on making assumptions regarding possible solute atom configurations. Although computing power has significantly improved over the past decades, exploring a sufficiently large number of cluster configurations such as is required to obtain the representative energetics of the system remains a very costly computation. Even if the investigation of each possible configuration was achievable, many are unlikely to be physically stable and would not be observed experimentally.

Atom probe tomography (APT) is a microscopy and microanalysis technique that maps in three-dimensions the distribution of atoms in a small volume of a solid material [16]. Given that APT can resolve atoms with sub-nanometre spatial resolution, it has strong potential for direct correlations with atomistic simulations. Although this potential has been discussed in some detail[17], it has not been widely harnessed. Notable efforts include kinetic Monte-Carlo (KMC) simulations [18–22], molecular dynamics [23] and density functional theory (DFT) [24,25]. In each of these examples, the results from comparison between the atom probe experiments and the computational simulations were qualitative at best, primarily due to difficulties in comparing the two data types. One the one hand, the simulation data produces 3D real space supercells with atoms assigned onto a rigid lattice, subject to the assumption-set of the calculation. On the other, spatial resolution limitations of APT result in 3D real space atomic-scale tomograms in which the atoms have a slight offset from their true average lattice positions [26]. The efficiency (< 80%) of detection of atoms in APT experiments is also a contributor to this complexity [27].

Recently, we proposed to use certain structural information contained in the atom probe data [28] to drive atoms back onto their most likely position within the original lattice [29]. In a second step, making use of atomic correlations extracted from the data [30], a new type of 'hybrid data' was generated [31]. The data-hybrid is corrected for the effect of detection efficiency, and is hence 100% complete. Our goal here was to explore the use of such data-hybrids for the purposes of comparison to and synergy with atomistic simulations. We use the hybrid atom probe data to demonstrate that only a limited number of atomic clusters configurations are encountered experimentally. These may be precisely characterised and subsequently used as the direct inputs for DFT calculations of (e.g.) the formation energy of such clusters as one proxy for their thermodynamic stability. Finally, the new capabilities offered by this approach are discussed.

In this work, hybrid atom probe data was generated from the analysis of an Al-1.1Cu-1.7Mg (at. %) alloy. This well-studied alloy was selected for this work because it exhibits a rapid increase (~60%) in hardness within the first 60 sec of thermal ageing at 150 °C following solution treatment [32]. Following detailed atomic resolution microscopy studies, including extensive atom probe microscopy, the effect was attributed to the fine and uniform dispersion of Cu-Mg co-clusters that form extremely rapidly upon elevated temperature ageing [32]. Termed 'cluster hardening', the effect was described as the result of an extremely fine dispersion of shearable obstacles, experimentally observed, leading to an increase in the dislocation friction under applied loads. The novelty of this finding, and the particular prominence given to the role of solute atom clusters attracted several follow-up studies, and subsequent investigations using calorimetry[33], independent atom probe studies[34,35], X-ray absorption spectroscopy (XAS)[36], positron annihilation spectroscopy (PAS) studies[35,37], and small-angle X-ray studies[38] have all confirmed the initial findings around the sequence of clustering, and the nexus between solute clustering and alloy hardening/strengthening.

Specimens were prepared from a model alloy of nominal composition that was prepared from elemental components of high purity (>99.99%) by induction melting and chill casting into a cast iron mould. The ingot was homogenized at 500 °C for 16 h, subsequently water quenched and cut into match-stick blanks with a square cross-section (~0.3 mm × 0.3 mm × 15 mm) to facilitate the specimen preparation for atom probe experiments. Sample blanks were solution treated for 1 h at 525 °C in a salt bath, cold water quenched and immediately aged in an oil bath for 60 seconds at 150 °C. Following heat treatment the blanks were sharpened into needle-like specimens suitable for atom probe (tip radius ∼50–100 nm) using a standard two-stage electropolishing technique, details contained elsewhere [39].

APT experiments were performed using a Cameca LEAP 3000X Si, using voltage pulsing mode at a repetition rate of 200 kHz, a pulse fraction of 25% and a set-point temperature of $20 \pm 5$ K. Data reconstruction was first performed using state-of-the-art algorithms [40] and calibration methods [41], and in a second step rectified using the methods and algorithms described in refs. [29,31]. Clusters detected in conventional APT data may actually represent the imaging of a significantly larger cluster of unknown size [42]. Atoms missing because of the limited detection efficiency can significantly affect characterization of nanostructure morphologies and hence measurements may not be truly indicative of the true shape. However, in the case of the hybrid data, the MC simulation [30,31] produces a clustered state having a known atomic configuration as defined by the generalised multicomponent short-range order (GM-SRO) parameter measured from the experimental atom probe data, such that the cluster size distribution and morphologies are representative of that within the original material. Visual comparison of both the experimental APT data and the hybrid APT data is shown in Figure 1.

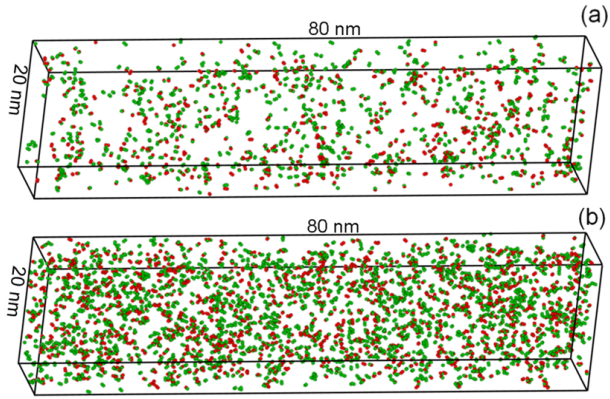

*Figure 1: Solute clusters in Al-1.1Cu-1.7Mg (at.%) aged for 60 seconds at 150 °C, displayed within 10×20×80 nm sub-volumes of (a) experimental APT data and (b) hybrid APT data after lattice rectification and MC simulation. Red spheres represent Cu atoms and green spheres represent Mg atoms, Al atoms are not shown for clarity.*

The position and composition of the clusters within the data was derived from applying a cluster-finding algorithm [5] which assumes that atoms belong to a cluster provided that they are separated by a distance smaller than a specific threshold [43]. Here, only when solutes are first nearest-neighbours to one another on the face-centred-cubic (FCC) Al-lattice, were they considered as belonging to a cluster. Al atoms were not considered to be part of the cluster. Over 5000 clusters were detected in this dataset. A relative frequency distribution of clusters containing 4 – 10 solute atoms as a function of their Mg concentration is shown in Figure 2 (a). Here we only refer to the composition in solute atoms, namely Mg and Cu, and the clusters can hence be described as $Mg_nCu_m$ with the cluster size being $(m+n)$. This distribution highlights that clusters are mostly Mg-rich, consistent with the literature[38]. Subsequently, for each solute cluster, a best-fit ellipsoid was performed to extract the dimensions of the cluster [44], allowing for an estimation of its aspect ratio and oblateness and thereby its approximate shape [45]. Figure 2 (b) shows a scatter plot of the aspect ratio as a function of oblateness, where the size of each dot is proportional to the cluster size and the colour of the dot relates to its composition: red is Cu and green is Mg, with all the shades in between corresponding to mixed clusters. This plot reveals that the clusters have a tendency to adopt a spheroidal shape, and exhibit a high-degree of compactness across the range of cluster compositions and sizes. Though we are cautious about the notion of a strict morphology for atomic clusters involving small (< 50) numbers of atoms in a metallic lattice, we note that these findings are consistent with previous reports[38].

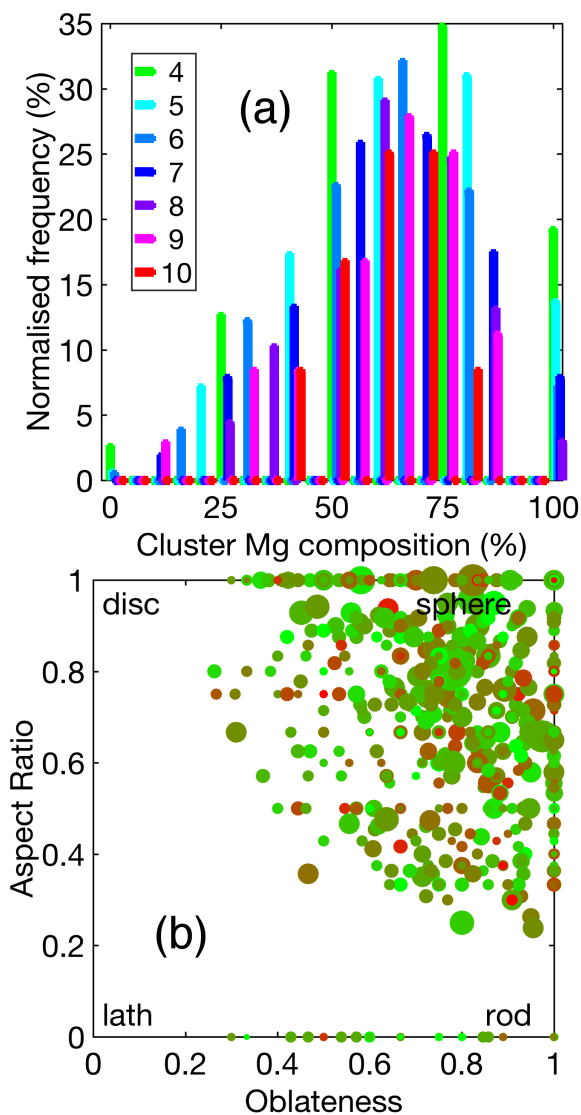

*Figure 2: (a) frequency vs. composition of clusters of m+n=4–10 solutes. (b) aspect ratio vs. oblateness for all clusters: the size of the dots is proportional to the size of the cluster, the colour is related to the cluster composition: red is 100% Cu while green is 100% Mg.*

To gain insights into the actual morphology and the energetics of these solute clusters, we performed extensive, all-electron DFT calculations, using the generalized gradient approximation [46] with the DMol$^3$ program package [47]. The wave functions were expanded in terms of a double-numerical quality basis set, with a set of large values of real-space cut-off (Cu: 10.99 Bohr; Mg: 12.09 Bohr; and Al: 12.75 Bohr). Here we confine our study to m+n = 4, 6 and 8 systems. At first, considering the large number of possible cluster configurations, we performed single-point energy calculations for the structures directly extracted from the APT hybrid data, to avoid the highly time-consuming relaxation step, as gas-phase clusters (i.e. surrounded by vacuum). This allowed for the computation of the total energy of the system for over 160 clusters. It became evident that the energetically favourable structures are close-packed isomers, so we therefore focused our attention on these for more detailed and precise calculations.

The position and elemental nature of individual solute atoms contained within a selection of 22 closely-packed clusters are shown in Figure 3. These clusters represent the most frequently encountered composition and combination of aspect ratio and oblateness. Each Mg-Cu solute cluster was embedded into a 108-atom FCC-Al based supercell. A reciprocal space of 2 x 2 x 2 K-point meshes was employed in the calculations of the 108-atom FCC-Al based supercells (i.e. Cu-Mg clusters embedded within an Al matrix). In this step, full atomic relaxation was allowed until the forces on the atoms were less than 0.005 eV/Å, in order to obtain precise energy measurements. To calculate the formation energy, metal-rich growth conditions were assumed, i.e., the atoms added to the FCC-Al are incorporated from the corresponding bulk reservoirs. Thus, the formation energy was calculated as $E_f = E_{Mg/Cu:Al} - E_{ref-Al} - n\,E_{Mg} - m\,E_{Cu} + (m+n)\,E_{Al}$, where $E_{Mg/Cu:Al}$, $E_{ref-Al}$, $E_{Mg}$, $E_{Cu}$ and $E_{Al}$ are the total energy of the Mg/Cu-Al supercell, the corresponding pure/undoped supercell, bulk hexagonal Mg, and FCC Cu and Al, respectively.

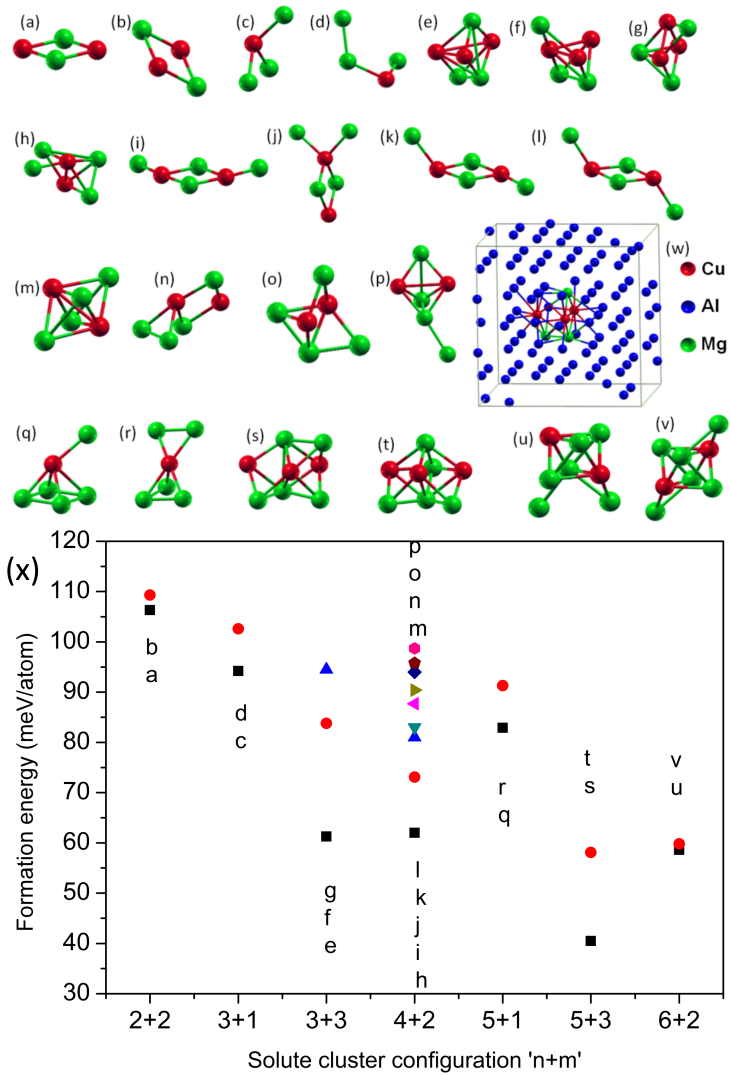

*Figure 3: Atomic geometry of the energetically favourable* m+n =4, 6 and 8 embedded cluster configurations. *The matrix Al atoms are omitted for clarity except for (w), which shows the actual structure of (e) embedded in the 108-atom cell; (x) Formation energy in meV/atom for a range of solute cluster configurations.*

Additionally, we have manually constructed a large number of presumably energetically favourable clusters (e.g. closely packed geometries), and performed similar calculations. This testing of additional geometries enabled assessment of the results with respect to potential shortcomings from the methodology of rectification and MC simulation to create the hybrid atom probe data. Nevertheless, the additional testing did not result in any new configuration with lower energy than the corresponding favourable ones from the hybrid data.

The formation energies of the clusters are shown in Figure 3 (x). These values agree rather well with those reported by Kovarik and Mills [48], who focused on the early stages of formation of Guinier–Preston–Bagaryatsky zones, but did not investigate similar structures and compositions. It is noteworthy, however, that we cannot directly compare our energy values with those determined by Kovarik and Mills (see Fig. 5a in ref. [48]), which refer to that of a structural transition energy, whereas we are reporting the ground state energies.

Our DFT results indicate that closely packed clusters are energetically favourable, which is especially evident for the 3+3 and 4+2 configurations shown in Figure 3 (e) and (h) are the most stable. This confirms the experimental results shown in Figure 2 (b). Interestingly, another important finding is that our simulation results suggest that Cu atoms prefer to locate in the middle, rather than the edge, of these Mg-Cu clusters. Comparing the two compact 4+2 configurations in Figure 3 (h) and (m), the former, which has two Cu atoms surrounded by four Mg atoms, is lower in energy than the later counterpart by 0.17 eV.

The oblateness and aspect ratios of the 4+2 cluster configurations shown in Figure 3 (h-p) were then extracted and superimposed with the experimental 4+2 clusters from Figure 3 (b) to create a new scatter plot shown in **Error! Reference source not found.**. The results from the clusters in both the experimental and hybrid data agree well. Minor discrepancies in morphology can be explained by the relaxation process in DFT that induces slight changes to the cluster shape, which were not included in the hybrid data. In addition, DFT calculations are purely ground-state energetics at a temperature of absolute zero, whereas the structures in the hybrid data include the effects from thermodynamics and kinetics. Interestingly, Figure 3 (x) shows that the formation energies predicted by DFT for the larger m+n=8 clusters (Figure 3 (s-v)) are consistently smaller by at least 20% than those for m+n=4 (Figure 3 (a-d)) and m+n=6 (Figure 3 (e-r)) clusters. The smaller clusters are experimentally more abundant suggesting that the evolution of the state of clustering follows discrete steps and must overcome particular energy barriers in order to stabilise progressively larger clusters.

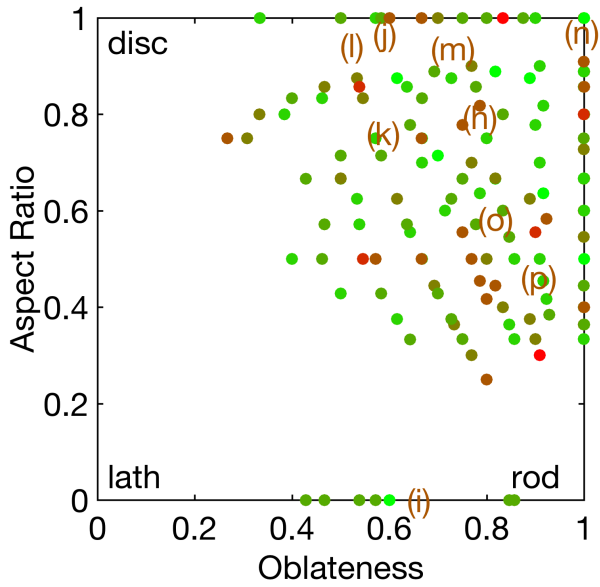

*Figure 4: Scatter plot of the aspect ratio vs. oblateness for the 4+2 clusters from the hybrid atom probe data, superimposed with those from the relaxed, calculated structure from DFT, (h)-(p).*

Finally, there is a vast body of literature which indicates that there can exist strong interactions between certain solutes and vacancies in Al [11,39,49,50]. Moreover, in the present Al-Cu-Mg alloys, it has been suggested that the vacancies are an intrinsic component of the clusters, contributing critically to their stability[32]. To investigate this, we introduced an Al-vacancy around, and far away from, selected clusters from Figure 3 (a, e, h, and s). Placing the Al-vacancy far from the Cu-Mg cluster within the supercell leads to a significant increase (i.e. ~ 0.3-0.6 eV) in the total energy than when the vacancy is in the vicinity of the Cu-Mg clusters, indicating that Al-vacancies will preferentially be found close to the cluster. Interestingly, our assessment of the influence of Al-vacancies using the hybrid data model suggests that the interaction between Cu and vacancies, as well as that between Mg and vacancies, are weak in Al and this is consistent with the ab initio calculations of Wolverton[49]. For instance, in the simple high-symmetry configuration denoted as (a) in Figure 3, the Al-vacancy prefers to reside above the rhombic cluster, which leads to an energy ~ 0.24 eV lower than the configurations of Al-vacancy bounded to single Cu or Mg atoms. Note that our calculated formation energy of an isolated Al vacancy is 0.58 eV, in good agreement with previous DFT results [51]. Therefore, our experimentally-informed DFT calculations suggest that Cu-Mg co-clustering is the resultant balance of attractive interactions between vacancies and Mg-Mg, Mg-Cu/Cu-Mg and Cu-Cu in the Al matrix. Consistent with other studies, the Al-vacancies are attracted to the vicinity of Cu-Mg clusters [49,35], and this follows since they are fundamental to the mechanism of solute atom diffusion that leads to cluster formation.

Deschamps et al.[38] reported that the energy required for a dislocation to shear a solute cluster less than 1 nm in diameter was approximately 0.5eV. It is tempting to compare this estimation to the energies of cluster formation, which range from 0.04-0.11 eV/solute atom, as calculated here. While we do consider that these quantities are related, we feel that they may not be directly compared. Beyond possible issues with the particular application of Freidel's model for strengthening from weak point-like obstacles[52] by Deschamps et al., which we consider is unlikely to capture the complexity of the dislocation-cluster interaction physics, other factors may contribute to expected differences between these types of energy values. The smaller size of the clusters observed here must play a key role in such comparisons, since only a fraction of the atomic environments would be modified upon interaction with a dislocation. We note also that our work may be an underestimate due to the aforementioned potential for vacancies to stabilise the cluster structures. In addition, we consider it unlikely that all clusters contribute to strengthening equally, and this may well lead to the need for complex distribution functions.

In conclusion, we have demonstrated how the APT hybrid data format introduced in ref. [31] enables a direct bridge with atomistic simulations (in this case, DFT) and hence allows direct coupling of microstructure and properties down to the atomic-scale. For such small clusters (e.g. up to 8 solute atoms), full-electron DFT is well-suited for this purpose and allowed us to derive precisely the morphology and associated formation energy of clusters that are encountered experimentally. In the rapid-hardening Al-Cu-Mg system, we have determined that compact, spheroidal clusters are more energetically favourable across a range of cluster sizes and compositions, in agreement with previous studies[38,39] An array of simulation techniques could later be deployed on APT hybrid data containing larger microstructural features (e.g. precipitates or interfaces), including large-scale orbital-free DFT [53], phase field crystal [54,55], molecular dynamics [56,57] or kinetic Monte-Carlo [58,59]. Some of these techniques could even find application in future developments of the methodology to generate the hybrid data. In the current work, DFT calculations allow precise measurement of the local, atomic configurational energy within the material, together with determination of the binding energies or interaction potential of each single atom within the dataset, paving the way for atomistic simulations with unparalleled precision at lower computational cost.

**Acknowledgements**

The authors acknowledge funding as well as scientific and technical input from the Australian Microscopy & Microanalysis Research Facility (AMMRF) at The University of Sydney. This research was undertaken with the assistance of resources from the National Computational Infrastructure (NCI), which is supported by the Australian Government. L.T. Stephenson is thanked for fruitful discussions in this topic over the years. BG is grateful for the support provided by Prof. Raabe and the group at MPIE.

**References**

[1] S.P. Ringer, G.C. Quan, and T. Sakurai, Mater. Sci. Eng. A **250**, 120 (1998).

[2] S.L. Shrestha, C. Zhu, G. Proust, F. Barbaro, C.R. Killmore, K. Carpenter, H. Kaul, K.Y. Xie, S.P. Ringer, and J.M. Cairney, Mater. Sci. Forum **753**, 559 (2013).

[3] S.P. Ringer, in *Mater. Sci. Forum* (2006), pp. 25–34.

[4] E. Meslin, B. Radiguet, P. Pareige, C. Toffolon, and A. Barbu, Exp. Mech. **51**, 1453 (2011).

[5] E.A. Marquis and J.M. Hyde, Mater. Sci. Eng. R Reports **69**, 37 (2010).

[6] M.J. Starink *, N. Gao, L. Davin, J. Yan, and A. Cerezo, Philos. Mag. **85**, 1395 (2005).

[7] T. Philippe, S. Duguay, D. Mathiot, and D. Blavette, J. Appl. Phys. **109**, (2011).

[8] K. Thompson, J.H. Booske, D.J. Larson, and T.F. Kelly, Appl. Phys. Lett. **87**, 52108 (2005).

[9] T. Homma, M.P. Moody, D.W. Saxey, and S.P. Ringer, Metall. Mater. Trans. A **DOI: 10.10**, (2012).

[10] L.T. Stephenson, M.P. Moody, and S.P. Ringer, Philos. Mag. **91**, 2200 (2011).

[11] V. Vaithyanathan, C. Wolverton, and L.Q. Chen, Acta Mater. **52**, 2973 (2004).

[12] A. Seko, Y. Koyama, and I. Tanaka, Phys. Rev. B **80**, 165122 (2009).

[13] M. Asta, C. Wolverton, D. de Fontaine, and H. Dreyssé, Phys. Rev. B **44**, 4907 (1991).

[14] C. Wolverton, M. Asta, H. Dreyssé, and D. de Fontaine, Phys. Rev. B **44**, 4914 (1991).

[15] C. Wolverton and D. de Fontaine, Phys. Rev. B **49**, 8627 (1994).

[16] T.F. Kelly and D.J. Larson, Annu. Rev. Mater. Res. **42**, 1 (2012).

[17] A. V. Ceguerra, A.J. Breen, L.T. Stephenson, P.J. Felfer, V.J. Araullo-Peters, P. V. Liddicoat, X. Cui, L. Yao, D. Haley, M.P. Moody, B. Gault, J.M. Cairney, and S.P. Ringer, Curr. Opin. Solid State Mater. Sci. **17**, 224 (2013).

[18] Z.G. Mao, C.K. Sudbrack, K.E. Yoon, G. Martin, and D.N. Seidman, Nat. Mater. **6**, 210 (2007).

[19] C. Pareige, F. Soisson, G. Martin, and D. Blavette, Acta Mater. **47**, 1889 (1999).

[20] I. Mouton, E. Talbot, C. Pareige, R. Lardé, and D. Blavette, J. Appl. Phys. **115**, 53515 (2014).

[21] G. Sha and A. Cerezo, Acta Mater. **53**, 907 (2005).

[22] E. Clouet, L. Lae, T. Epicier, W. Lefebvre, M. Nastar, and A. Deschamps, Nat. Mater. **5**, 482 (2006).

[23] X.W. Zhou, H.N.G. Wadley, R.A. Johnson, D.J. Larson, N. Tabat, A. Cerezo, A.K. Petford-Long, G.D.W. Smith, P.H. Clifton, R.L. Martens, and T.F. Kelly, Acta Mater. **49**, 4005 (2001).

[24] B. Gault, X.Y. Cui, M.P. Moody, F. De Geuser, C. Sigli, S.P. Ringer, and A. Deschamps, Scr. Mater. **66**, 903 (2012).

[25] Y. Amouyal, Z. Mao, and D.N. Seidman, Acta Mater. **74**, 296 (2014).

[26] B. Gault, M.P. Moody, F. De Geuser, A. La Fontaine, L.T. Stephenson, D. Haley, S.P. Ringer, A. La Fontaine, L.T. Stephenson, D. Haley, and S.P. Ringer, Microsc. Microanal. **16**, 99 (2010).

[27] G. Da Costa, F. Vurpillot, A. Bostel, M. Bouet, and B. Deconihout, Rev. Sci. Instrum. **76**, 13304 (2005).

[28] B. Gault, M.P. Moody, J.M. Cairney, and S.P. Ringer, Mater. Today **15**, 378 (2012).

[29] M.P. Moody, B. Gault, L.T. Stephenson, R.K.W. Marceau, R.C. Powles, A. V Ceguerra, A.J. Breen, and S.P. Ringer, Microsc. Microanal. **17**, 226 (2011).

[30] A. V Ceguerra, M.P. Moody, R.C. Powles, T.C. Petersen, R.K.W. Marceau, and S.P. Ringer, Acta Crystallogr. Sect. A **68**, (2012).

[31] M.P. Moody, A. V Ceguerra, A.J. Breen, X.Y. Cui, B. Gault, L.T. Stephenson, R.K.W. Marceau, R.C. Powles, and S.P. Ringer, Nat. Commun. **5**, 5501 (2014).

[32] S.P. Ringer, T. Sakurai, and I.J. Polmear, Acta Mater. **45**, 3731 (1997).

[33] M.J. Starink, A. Cerezo, J.L. Yan, and N. Gao, Philos. Mag. Lett. **86**, 243 (2006).

[34] M.J. Starink *, N. Gao, L. Davin, J. Yan, and A. Cerezo, Philos. Mag. **85**, 1395 (2005).

[35] R.K.W. Marceau, G. Sha, R. Ferragut, A. Dupasquier, and S.P. Ringer, Acta Mater. **58**, 4923 (2010).

[36] T.E.M. Staab, M. Haaks, and H. Modrow, Appl. Surf. Sci. **255**, (2008).

[37] B. Klobes, T.E.M. Staab, M. Haaks, K. Maierand, and I. Wieler, Phys. Status Solidi - Rapid Res. Lett. **2**, (2008).

[38] A. Deschamps, T.J. Bastow, F. de Geuser, A.J. Hill, and C.R. Hutchinson, Acta Mater. **59**, 2918 (2011).

[39] R.K.W. Marceau, G. Sha, R.N. Lumley, and S.P. Ringer, Acta Mater. **58**, 1795 (2010).

[40] B. Gault, D. Haley, F. de Geuser, M.P. Moody, E.A. Marquis, D.J. Larson, and B.P. Geiser, Ultramicroscopy **111**, 448 (2011).

[41] B. Gault, M.P. Moody, F. de Geuser, G. Tsafnat, A. La Fontaine, L.T. Stephenson, D. Haley, and S.P. Ringer, J. Appl. Phys. **105**, 34913 (2009).

[42] L.T. Stephenson, M.P. Moody, B. Gault, and S.P. Ringer, Microsc. Res. Tech. **74**, 799 (2011).

[43] J.M. Hyde and C.A. English, in *MRS 2000 Fall Meet. Symp.*, edited by Lucas RGE, Snead L, Kirk MAJ, and Elliman RG (Boston, MA, 2001), pp. 27–29.

[44] R.A. Karnesky, C.K. Sudbrack, and D.N. Seidman, Scr. Mater. **57**, 353 (2007).

[45] R.K.W. Marceau, L.T. Stephenson, C.R. Hutchinson, and S.P. Ringer, Ultramicroscopy **111**, 738 (2011).

[46] J.P. Perdew, K. Burke, and M. Ernzerhof, Phys. Rev. Lett. **78**, 1396 (1997).

[47] B. Delley, J. Chem. Phys. **113**, 7756 (2000).

[48] L. Kovarik and M.J. Mills, Acta Mater. **60**, 3861 (2012).

[49] C. Wolverton, Acta Mater. **55**, 5867 (2007).

[50] C.R. Hutchinson, B.M. Gable, N. Ciccosillo, P.T. Loo, T.J. Bastow, and A.J. Hill, in *Alum.*

*Alloy. Their Phys. Mech. Prop. Vol. 1*, edited by J. Hirsch, B. Strotzki, and G. Gottstein (Wiley-VCH Verlag GmbH & Co. KGaA, Weinheim, 2008), pp. 788–794.

[51] K. Carling, G. Wahnström, T.R. Mattsson, A.E. Mattsson, N. Sandberg, and G. Grimvall, Phys. Rev. Lett. **85**, 3862 (2000).

[52] J. Friedel, *Dislocations* (Pergamon Press, Oxford, 1964).

[53] V. Gavini, K. Bhattacharya, and M. Ortiz, J. Mech. Phys. Solids **55**, 697 (2007).

[54] M. Greenwood, J. Rottler, and N. Provatas, Phys. Rev. E **83**, 31601 (2011).

[55] J. Berry, N. Provatas, J. Rottler, and C.W. Sinclair, Phys. Rev. B **89**, 214117 (2014).

[56] F. Ulomek and V. Mohles, Acta Mater. (2016).

[57] B.J. Alder and T.E. Wainwright, J. Chem. Phys. **31**, 459 (1959).

[58] A.T. Wicaksono, C.W. Sinclair, and M. Militzer, Model. Simul. Mater. Sci. Eng. **21**, 85010 (2013).

[59] P.E. Goins and E.A. Holm, Comput. Mater. Sci. **124**, 411 (2016).